\documentclass[12pt,titlepage]{article}
\usepackage[english]{babel}
\usepackage[dvips]{graphicx}

\begin{document}
\begin{titlepage}

\begin{center}
{\Large \bf   CP violation  in  the partial width 
asymmetries for   $B^- \to \pi^+ \pi^- K^-$
and  $B^- \to K^+ K^- K^-$    decays  }\\
\vspace{1cm}
{\large \bf S. Fajfer$^{a,b}$,  T.N. Pham$^{c}$ and A. Prapotnik$^{a}$\\}

{\it a) J. Stefan Institute, Jamova 39, P. O. Box 3000, 1001 Ljubljana, Slovenia}
\vspace{.5cm}

{\it b)  Department of Physics, University of Ljubljana, Jadranska 19,
1000 Ljubljana, Slovenia}
\vspace{.5cm}

{\it c) Centre de Physique Theorique, Centre National de la Recherche Scientifique,
UMR 7644, Ecole Polytechnique, 91128 Palaiseau Cedex, France\\}

\end{center}

\vspace{0.25cm}

\centerline{\large \bf ABSTRACT}

\vspace{0.2cm}

We investigate a possibility of observing  CP asymmetries in the
partial widths for the decays $B^- \to \pi^+ \pi^- K^-$ and $B^- \to
K^+ K^- K^-$  produced by  the interference of the non-resonant
decay amplitude with the resonant amplitudes. The resonant states
which subsequently decay into $\pi^+ \pi^- $ and $K^+ K^- $ or $K^-
\pi^+$ are charmonium $\bar c c$ states with $J^P = 0^+, 1^-, 1^+$ 
or the $\phi(1020)$ meson. We find that the largest partial
width asymmetry comes from the $\chi_{c0}$ resonance, while the
resonance $\psi(2S)$ gives a partial width asymmetry of the order
$10\%$.

   \end{titlepage}

\section{INTRODUCTION}

The experimental data on $B$ mesons decays into three mesons
accumulate \cite{fry} - \cite{BaBar} and a number of 
important questions on their decay dynamics and their 
relevance for  the precise determination of the CP violating phase $\gamma$ 
should be answered \cite{FOP} - \cite{Hazumi}. Motivated by Belle and 
BaBar results 
on the $B$ mesons three-body decays \cite{fry,Belle1,Belle2,BaBar}, 
we continue with the study of CP violating partial width 
asymmetry in the $B^\pm \to K^\pm \pi^+ \pi^-$ and 
$B^\pm \to K^\pm K^+ K^-$ decay amplitudes.

Recently, we have studied a case of the partial width asymmetry
resulting from the interference of the non-resonant $B^- \to M^+ M^-
K^-$, $M=\pi, K$, and the resonant $B^- \to \chi_{c0} K^- \to M^+
M^-K^-$ decay amplitudes \cite{FOP}.  In both decay modes, the
dominant contribution to the non-resonant amplitude comes from the
penguin operators. However, there is a small
tree level contribution in which enters the weak CP violating phase
$\gamma$. The strong phase, which is necessary to obtain 
the CP violating asymmetry,  enters trough the dispersive part of 
both non-resonant
and resonant amplitudes. 

It was pointed out  by the authors of \cite{EGM} and 
\cite{SEW} that the dispersive part of the non-resonant amplitude 
exactly cancels the dispersive part of the resonant amplitude
coming from the intermediate state which is  
identical to the final state. Therefore,
the partial width asymmetry for $B^\pm \to {\cal R} K^\pm \to M^+ M^-
K^\pm$, $M=\pi, K$, will be proportional to the decay width of the resonant
state ${\cal R}$ to all channels excluding the  $M^+ M^-$ state. 
It means that one would expect a large
CP asymmetry for the two-meson invariant mass in the $\chi_{c0}$ mass
region since the decay width of $\chi_{c0}$ is rather large and its
branching ratio to $M^+ M^-$,
$M=\pi,K$ is negligible.  The amplitude for the $\chi_{c0}$ resonant
decay mode was determined using the narrow width approximation
\cite{FOP,DEHT} and the experimental results for the $B^- \to \chi_{c0} K^-$
and $\chi_{c0} \to M^+ M^-$ decay rates. The asymmetry was found to be
about $20\%$. In the case of $B^- \to K^- M^+ M^-$ there are,  
however,  additional important reasons why the partial width asymmetry 
can be sizable. In fact, if in the $B^- \to K^- M^+ M^-$ decays the partial widths coming 
from the  non-resonant  ${\cal M}_{\rm nr}$  and the resonant 
 ${\cal M}_{\rm r}$ amplitude are of the same order of magnitude, 
 as in our analysis 
at the $\chi_{c0}$  resonance region \cite{FOP}, one obtains  a significant CP violating asymmetry. 
In the case of  negligible  non-resonant amplitude relative to the resonant
amplitude (or vice versa) one would get a  very small partial width asymmetry. 

In this paper, we extend this  analysis to  the  case of the 
CP violating partial width asymmetry when the interference with  
the non-resonant amplitude occurs  
in the neighborhood of  the resonance  ${\cal R}$ which is  a charmonium 
$\bar c c$ state with $J^P = 0^+, 1^-, 1^+$ or a light vector and scalar meson.
We will restrict our investigation only to  those resonant states ${\cal R}$
for which the decay $B^- \to {\cal R} M^-$, $M=K, \pi$ amplitude does
not have two or more contributions with different weak phase, as 
from the experimental branching ratio we are able to extract only the
absolute value of the amplitude. 
For example in the case of $B^- \to {\cal R} K^-$ with  ${\cal R} =
\rho^0$ there is a penguin and a tree  
amplitude and one needs to know their relative sizes to constrain the partial width asymmetry.
In this decay mode it has also been found that the naive factorization fails 
to describe the decay rate \cite{heff1,heff2}.
Therefore, we concentrate on the partial width asymmetry for the 
cases in which the relevant two-body amplitude can be completely extracted from the  
measured decay rates.

In the case of the $B^- \to K^- \pi^+ \pi^-$ partial width asymmetry, 
the intermediate resonant states of interest  would be 
the light strange mesons $K^*(890)$, $K_1(1270)$, $K_1(1400)$,
$K^{*}_{0}(1430)$ etc. in the decay chain 
$B^- \to {\cal R} \pi^- \to K^-
\pi^+ \pi^-$  and the  charmonium $\bar cc$ states in the decay chain $B^-
\to {\cal R} K^- \to K^- \pi^+ \pi^-$. 
The $B^-$ decays to these strange mesons 
 in the final state occur as  a pure penguin transition.
Among  all such decays
only the rates for ${\cal R}=K^*(890)$ and
$K^{*}_{0}(1430)$ were measured \cite{Belle2}. 
 However, the   
$K^*(890)$ and $K^{*}_{0}(1430)$ mesons decay  to
$K^- \pi^+$ with the branching ratios  close to 
$100\%$. In the case which we consider  it means that the 
partial decay width to the rest of the states 
is negligible and the corresponding 
CP violating asymmetry vanishes.
The relevant charmonium $\bar cc$ states in the decay chain $B^-
\to {\cal R} K^- \to \pi^+ \pi^- K^-$ are produced by the 
$b \to \bar c c s $ transition.  
The resonant $B^- \to M^{+}M^{-} K^-$ amplitude is obtained from 
the tree level contribution which is proportional to the 
$V_{cb}$ and $V_{cs}$ CKM matrix elements, followed by  the  strong
decay of the $\bar cc$ state  into $\pi^+ \pi^-$  or  $K^+ K^-$  
via the OZI (Okubo-Zweig-Iizuka) suppressed strong
interaction. Apart from the already mentioned $\chi_{c0}$ state, this
category  includes also $J/\psi$, $\chi_{c1}$, $\chi_{c2}$,
$\psi(2S)$ etc. We will consider contributions from all 
the above mentioned states,  even though the $B^- \to \chi_{c2} K^-$
and $\chi_{c1} \to M^-M^+$ branching ratios have  not been measured yet. 
Nevertheless, we
expect that the partial width asymmetry in this decay modes can be
rather large.  Although one would expect that the $b \to \bar c c s $ 
transition will 
give larger rates for the two-body decays than in the case of the 
$b \to \bar u u s$ transition, the fact that the 
strong transition of the charmonium states is  OZI suppressed
makes the non-resonant 
and resonant partial width  to be of 
the same size and this leads to a sizable CP violating asymmetry. 

In the case of the $B^- \to K^- K^+ K^-$ decays with the two-meson invariant mass 
below the charmonium production threshold, the resonant contribution
comes from the intermediate  $\bar ss$ states. We consider only  
the CP asymmetry at the $\phi(1020)$ resonance  
and do not consider contributions from  
the scalar meson resonances
due to  the  lack of knowledge on their  structure. 

In the analysis of the  partial width CP asymmetry,  one needs a   
knowledge of the non-resonant amplitudes. 
We compute the non-resonant decay amplitudes by using a 
model which combines the heavy quark effective theory and chiral Lagrangian, 
previously developed in   \cite{FOP} - \cite{Cheng}.  This model assumes the naive 
factorization for the weak vertices. 
The fact that the factorization works reasonably well in the relevant 
two-body decay modes encourages us to apply
it in the three-body decays we consider here. Even more, the experimental investigation  of  
the non-resonant amplitudes done by Belle collaboration \cite{Belle2} indicates that one has 
to  rely on a model when discussing 
the non-resonant background.
In comparison with our previous investigation \cite{FOP,BFOPP}, we include now the  contributions of 
$B^{*}(0^+)$ resonances. 

In Section 2 we present the calculation and the results on the non-resonant $B^- \to K^-
M^+ M^-$, $M=\pi,K$ decay modes, while in Section 3 we analyze the 
partial width asymmetries. The summary of our results is given in 
Section 4.

\vspace{1cm}

\section{ NON-RESONANT AMPLITUDES}
 
The effective weak Hamiltonian relevant for the 
$B^\pm \to K^\pm M^+ M^-$ decays and their $CP$ conjugates 
after Fierz reordering of the quark fields and neglecting the contribution 
of the color octet operators is
\cite{heff1} - \cite{heff6}:
\begin{equation}
{\cal H}_{eff}=\frac{G_f}{\sqrt 2}(V_{us}^*V_{ub}
(a_1 O_1+a_2 O_2)+V_{cs}^*V_{cb}(a_{1c} O_{1c}+a_{2c} O_{2c})-V^*_{ts}V_{tb}\sum_{i=3}^{10} a_i O_i)\,,
\end{equation}
The effective Wilson coefficients are denoted by $a_i$ 
and the operators $O_i$  read:
\begin{eqnarray}
O_1=(\bar ub)_{V-A}(\bar su)_{V-A}\,,
\qquad
O_2=(\bar uu)_{V-A}(\bar sb)_{V-A}\,,\\ 
O_{1c}=(\bar cb)_{V-A}(\bar sc)_{V-A}\,,
\qquad
O_{2c}=(\bar cc)_{V-A}(\bar sb)_{V-A}\,,\\ 
O_3=\sum_{q=u,d,s}  (\bar qq)_{V-A}(\bar sb)_{V-A}\,,
\qquad
O_4=\sum_{q=u,d,s} (\bar qb)_{V-A}(\bar sq)_{V-A}\,,\\
O_5=\sum_{q=u,d,s} (\bar qq)_{V+A}(\bar sb)_{V-A}\,,
\qquad 
O_6=-2\sum_{q=u,d,s} (\bar q(1-\gamma_5)b)(\bar s(1+\gamma_5)q)\,,\\
O_7=\sum_{q=u,d,s}\frac{3}{2}e_q (\bar qq)_{V+A}(\bar sb)_{V-A}\,,
\qquad
O_8=-2\sum_{q=u,d,s}\frac{3}{2}e_q (\bar q(1-\gamma_5)b)
(\bar s(1+\gamma_5)q)\,,\\
O_9=\sum_{q=u,d,s}\frac{3}{2}e_q  (\bar qq)_{V-A}(\bar sb)_{V-A}\,,
\qquad
O_{10}=\sum_{q=u,d,s}\frac{3}{2}e_q (\bar qb)_{V-A}(\bar sq)_{V-A}\,,
\end{eqnarray}
where $(\bar q_1 q_2)_{V \pm A}$ stands for   
$\bar q_1\gamma^\mu(1 \pm \gamma_5)q_2$. Here $O_1$ and $O_2$ are the tree level operators,
$O_3-O_6$ are the QCD penguin operators and $O_7-O_{10}$ are
the electromagnetic penguin operators. From \cite{heff6} we take
$a_1=1.05$, $a_2=0.07$, $a_4=-0.043-0.016i$
and $a_6=-0.054-0.016i$. The values of the other Wilson coefficients are at
least one order of magnitude smaller and therefore we can safely
neglect them.

For the  CKM matrix 
elements the Wolfenstein
parametrization is used ($V_{ub}=A \lambda^3 (\bar \rho-i \bar \eta)$,
$V_{us}=\lambda$, $V_{ts}=-A\lambda^2$, $V_{tb}=1$), with $A=0.8$,
$\lambda=0.228$, $\bar \rho=0.118-0.273$ (the average value 0.222) and
$\bar \eta=0.305-0.393$ (the average value 0.339) \cite{ckmfit}.
The matrix elements of the four quark operators 
acting in  $O_i$ for the 
$B^- \to K^- \pi^+ \pi^-$ decay can be written 
using the factorization assumption as:

\begin{equation}
\label{1}
<\pi^+ \pi^- K^-|(\bar s b)(\bar q q)|B^->=
<K^-|(\bar s b)|B^-><\pi^-\pi^+|(\bar q q)|0>\,,
\end{equation}
\begin{equation}
\label{2}
<\pi^+ \pi^- K^-|(\bar d b)(\bar s d)|B^->=$$$$
<\pi^-|(\bar d b)|B^-><K^- \pi^+|(\bar s d)|0>\,,
\end{equation}
\begin{equation}
\label{3}
<\pi^+ \pi^-  K^-|(\bar u b)(\bar s u)|B^->=
<\pi^+ \pi^-|(\bar u b)|B^-><K^-|(\bar s u)|0>
\end{equation}
$$+<0|(\bar u b)|B^-><K^-\pi^+\pi^-|(\bar s u)|0>\,.$$ In the above
equations $(\bar q_i q_j)$ denotes the vector or axial-vector
current or scalar or pseudoscalar density.  By analyzing
the matrix elements given  above, one finds \cite{FOP} that only the first
term in (\ref{3}) gives important contribution to  the non-resonant decay
rate.  Terms (\ref{1}) and (\ref{2}) contribute to the resonant
part of the amplitude (through resonances which decay into $\pi^+
\pi^-$ or $K^- \pi^-$ respectively), while the annihilation term in (\ref{3})
is found to be negligible as explained in \cite{FOP}. In the matrix element
of the $O_6$ operator, additional terms might arise, but they
are either small or cancel among themselves \cite{FOP}. 

The $B^- \to K^- K^+ K^-$ amplitude can be factorized in the same way
by replacing $\pi^\pm$ with $K^\pm$ in (\ref{1})-(\ref{3}). However,
in this case, the contribution coming from $B^- \to \rho^0 K^- \to K^-
K^+ K^-$ {(Eq. (\ref{1}))} is part of the non-resonant
amplitude, since the $\rho^0$ mass is below the $K^- K^+$ threshold. 
Nevertheless, we find this contribution to be small due to the 
suppression of the  $\rho^0$ propagator in the high energy regions and due
to the smallness of its Wilson coefficients ($a_2$ and $a_9$) and will
therefore neglect it. The same argument
holds if the $\rho$ meson is replaced by similar resonances 
($\sigma$ etc.).

\begin{figure}
\begin{center}
\includegraphics[width=15cm]{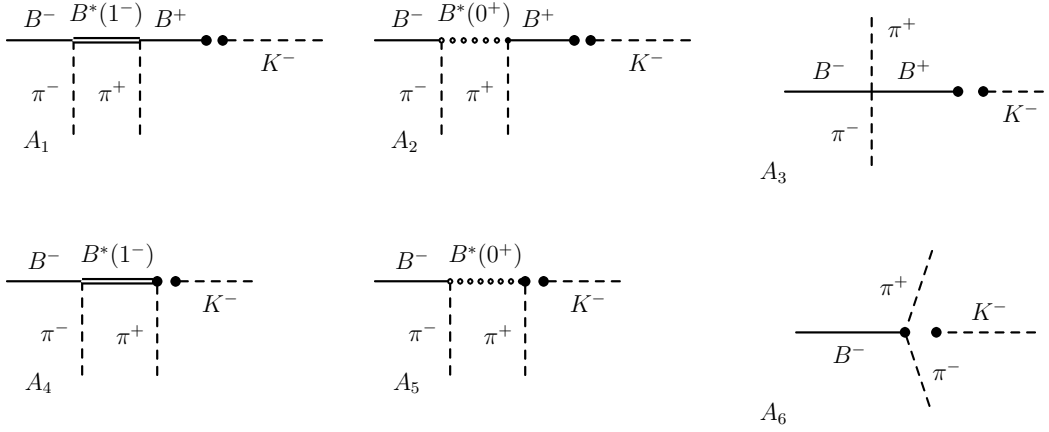}
\label{diagrami}
\caption{Feynman diagrams contributing to the non-resonant part of the
amplitude.}
\end{center}
\end{figure}

Next, we proceed with the determination of  
${\cal A}_\pi= \langle\pi^+(p_2)\pi^-(p_1)K^-(p_3)|O_1|\bar B^-\rangle$ and 
${\cal A}_K=$ $\langle K^+(p_2) K^-(p_1)  K^-(p_3)|O_1|\bar
B^-\rangle$. The  approach used in the calculation of 
these matrix elements was 
already explained 
in \cite{FOP,nrracun,beta,O6}. Here, we follow the same method,  
but 
add the contributions of the $B^{*}_{0}$ scalar meson resonances. We introduce 
the $b \bar q$ states ($q=u,d,s$),
with the $J^P=1^+,0^+$ assignment incorporated
in the field $S$ \cite{casa}:
\begin{equation}
S =\frac{1}{2}(1+v_\alpha\gamma^\alpha)[D_1^\mu\gamma_\mu\gamma_5-D_0],
\label{S1}
\end{equation}  
which then interacts with the $b\bar q$ $J^P= 1^-, 0^-$ multiplet ($H$)
and the pseudo Goldstone mesons by the means of the Lagrangian:
\begin{equation}
{\cal L}_s = i h Tr (S_b\gamma_\mu \gamma_5 {\cal A}^{\mu}_{ba} 
\bar H_a),
\label{S2}
\end{equation}
where ${\cal A}^{\mu}= 1/2$ ($ \xi^\dag \partial^\mu \xi - 
\xi \partial^\mu \xi^\dag)$ with the the light 
pseudoscalars fields in $\xi$. The weak current is given by
\begin{equation}
j_{\mu}^S = i\frac{F^+}{2} Tr [\gamma^\mu (1- \gamma_5) 
S_b \xi^\dag_{ba}]. 
\label{S3}
\end{equation}
The parameter  $h=-0.6\pm0.2$ 
is taken from the recent study of the $D_s (0^+)$ state  
\cite{Col}, 
while for the scalar meson decay constant we  use  
$F^+=0.46$ GeV$^{3/2}$ \cite{casa}. 

The matrix element ${\cal A}_\pi$ can be   
written as:

\begin{equation}
{\cal A}_\pi=-f_3[m_3^2 r^{\rm nr}+1/2(m_B^2-m_3^2-s) w_+^{\rm nr}+
1/2(s+2t-m_B^2-2m_2^2-m_3^2) w_-^{\rm nr})]\,,
\label{A}
\end{equation}
where the form factors $w_+^{\rm nr}$, $w_-^{\rm nr}$ and 
$r^{\rm nr}$ are 
determined by calculating contributions coming form the Feynman diagrams in Fig. 1:
\begin{equation}
w_+^{\rm nr}=-\frac{g}{f_1 f_2}\frac{f_{B^*}m_{B^*}^{3/2}m_B^{1/2}}
{t-m^2_{B^*}}\left(1-\frac{m_B^2-m_1^2-t}{2m^2_{B^*}}\right)+
\frac{f_B}{2f_1 f_2}
\end{equation}
$$
-\frac{\sqrt{m_B} \alpha_2}{2f_1 f_2 m_B^2}
(2t+s-m_B^2-m_3^2-2m_1^2)
+\frac{F^+ h \sqrt{m_B}}{2f_1 m^2_B}\frac{m_B^2-t}{t-m_{B^*_0}^2}\,,
$$
\begin{equation}
w_-^{\rm nr}=\frac{g}{f_1 f_2}\frac{f_{B}m_{B^*}^{3/2}m_B^{1/2}}
{t-m^2_{B^*}}\left(1+\frac{m_B^2-m_1^2-t}{2m^2_{B^*}}\right)+
\frac{\sqrt{m_B} \alpha_1}{f_1 f_2}+
\frac{F^+ h \sqrt{m_B}}{2f_1 m^2_B}\frac{m_B^2-t}{t-m_{B^*_0}^2}\,,
\end{equation}
\begin{equation}
r^{\rm nr}=-\frac{f_B}{2f_1 f_2(m_3^2-m_B^2)}(2t+s-m_B^2-m_3^2-2m_1^2)+
\frac{f_B}{2f_1 f_2}
\label{r}
\end{equation}
$$+\frac{gf_{B}}{f_1f_2(t-m_{B^*}^2)}(m_B^2-m_1^2-t)
-\frac{\sqrt{m_B} \alpha_2}{2f_1 f_2 m_B^2}(2t+s-m_B^2-m_3^2-2m_1^2)$$
$$-\frac{4g^2f_B m_{B} m_{B^*}}{f_1 f_2(m_3^2-m_B^2)(t-m_{B^*}^2)}
\left(\frac{s-m_1^2-m_2^2}{2}-\frac{(t+m_2^2-m_3^2)(m_B^2-m_1^2-t)}{4m_B^{*2}}
\right)$$
$$+\frac{F^+ h \sqrt{m_B}}{f_1 m^2_B}\frac{m_B^2-t}{t-m_{B^*_0}^2}
+\frac{F^+h^2\sqrt{m_B}}{f_1 f_2 m^3_B}\frac{(m_B^2-t)(t-m_3^2)}
{(t-m_{B^0}^2)(m_3^2-m_{B^*_0}^2)}\,.$$
We used the Mandelstam's variables $s=(p_B-p_3)^2$ and $t=(p_B-p_1)^2$. 
Indies 1, 2 and 3 correspond to $\pi^-$, $\pi^+$ and $K^-$
respectively ($f_1=f_2=f_\pi$, $f_3=f_K$,
$m_1=m_2=m_\pi$, $m_3=m_K$). The masses $m_B$, $m_{B^*}$ and
$m_{B^*_0}$ correspond to the $B^-$, $B^{0*}(1^-)$ and $B^{0*}(0^+)$
mesons, $(1^-)$ denoting vector and $(0^+)$ scalar states.
The rest of parameters are  taken to be $f_\pi=0.132\,$GeV, 
$f_K=0.16\,$GeV, $f_B=0.175\,$GeV, $f_{Bs}=1.16f_B$,
$\alpha_1=0.16\,$GeV$^{1/2}$, $\alpha_2=0.15\,$GeV$^{1/2}$ as in 
\cite{FOP}. For the strong coupling $g$ we use
$g=0.56$ according to the measurement of \cite{CLEO}. 
 Note that in \cite{FOP} there are misprints in 
Eq.(16): the sign in front of $\alpha_2$ is reversed, 
as well as the overall sign in (22). 

The matrix element of  $O_4$  
has the same structure as the matrix element of $O_1$ 
while for determining the matrix element of  $O_6$
we follow the approach described in \cite{FOP}. 
 Using the expressions  (18)-(20) of
\cite{FOP}, we find  that its contribution is
proportional to the matrix element of 
$O_1$ or $O_4$ with the proportionality  
factor $k_6=-2\frac{{\cal B}f_\pi^2}{m_b f_K^2}$. 

The  matrix element 
$\langle K^+(p_2) K^-(p_1) K^-(p_3)|O_1|\bar B^-\rangle$ 
is calculated in the same  way. The expression  for ${\cal
A}_K$ and its form factors can be derived from the Eqs. (\ref{A})-(\ref{r}),
adding the additional contribution obtained by  interchanging
$s$ and $t$ and by taking $f_1=f_2=f_3=f_K$, $m_1=m_2=m_3=m_K$. 
In the propagators the $B$ meson masses are replaced  by the
$B_s$ mass.

Now, the non-resonant amplitudes for  $B^- \to M^+ M^- K^-$ can be written as
\begin{equation}
{\cal M}_{\rm nr}=\frac{G_f}{\sqrt{2}}{\cal A}_{K,\pi}(V_{ub}V^*_{us}a_1-
V_{tb}V^*_{ts}(a_4+k_6 a_6))\,,
\end{equation}
with ${\cal A}_{\pi,K}$ defined in Eq. (\ref{A}-\ref{r}), while for
$B^+ \to M^+ M^- K^+$ we have:
\begin{equation}
\bar {\cal M}_{\rm nr}=\frac{G_f}{\sqrt{2}}{\cal A}_{K,\pi}(V^*_{ub}V_{us}a_1-
V^*_{tb}V_{ts}(a_4+k_6 a_6))\,.
\end{equation}
Using {the above expressions}, we obtain the following branching rations:
\begin{equation}
{\rm BR}(B^\pm \to K^\pm \pi^+ \pi^-)_{\rm nr}=9.0 \times 10^{-6}\,,
\qquad {\rm BR}(B^\pm \to K^\pm K^+ K^-)_{\rm nr}= 14 \times 10^{-6}\,,
\end{equation}
 where ${\rm BR}(B^\pm \to K^\pm M^+ M^-)_{\rm nr}$ stands for  the  
CP-averaged rates for $B^-\to M^- M^+ K^-$ and 
$B^+\to M^- M^+ K^+$ 
($({\rm BR}(B^- \to K^- M^+ M^-)+ {\rm BR}(B^+ \to K^+ M^+ M^-))/2$).
In \cite{FOP} it was found that  due to
the imaginary part of the $a_4$ and $a_6$ Wilson coefficients we 
can have a large CP asymmetry between  the non-resonant 
$B^+\to M^- M^+ K^+$ and $B^-\to M^- M^+
K^-$ amplitudes. The size of this asymmetry depends on 
the $\bar \rho$ and $\bar \eta$ CKM
parameters and is rather large ($60\%$ in the case of 
$B^\pm\to \pi^- \pi^+ K^\pm$
and $40\%$ in a case of $B^\pm\to K^- K^+ K^\pm$ decay mode). 
The largest error in 
${\rm BR}(B^\pm \to K^\pm M^+ M^-)_{\rm nr}$, due to the model
parameters, comes from the uncertainty in the CKM weak phase
$\gamma$, the decay constants and the coupling
$g$. For example, by taking two times smaller  $g$, 
the rate  ${\rm BR}(B^\pm \to K^\pm \pi^+ \pi^-)_{\rm nr}$ decreases by $40\%$ and 
${\rm BR}(B^\pm \to K^\pm \pi^+ \pi^-)_{\rm nr}$ by $30\%$. Varying $\bar \rho$
between 0.118 and 0.273 and $\bar \eta$ between 0.305 and 0.393 gives 
${\rm BR}(B^\pm \to K^\pm \pi^+ \pi^-)_{\rm nr}=(6.2-12.6)\times 10^{-6}$ and
${\rm BR}(B^\pm \to K^\pm K^+ K^-)_{\rm nr}= (11-17) \times 10^{-6}\,$. 
The uncertainty in the branching ratios coming from the $B$ decay
constants is not larger than $10\%$. 

The Dalitz plots for $B^- \to K^- M^+ M^-$ ($M= \pi, K$) decays, 
 are given
in Fig. 2 ($g=0.56$).
 We can see, that the non-resonant $B^- \to K^- K^+ K^-$ decay
amplitude is rather flat, while in the
case of $B^- \to \pi^- \pi^+ K^-$, an increase at low $K$ and $\pi$
momenta phase space region is evident. 
The inclusion of the scalar states $B^{*}(0^+)$ is not giving significant 
contribution to the decay rate,  increasing it by few 
percent in both decay modes.

Recently  B factories \cite{fry,Belle2,BaBar} 
got some insight into the 
nonresonant contribution to the $B^- \to K^- M^+ M^-$ decay widths. The preliminary results 
of the Belle collaboration are \cite{fry,Belle2}:
${\rm BR}(B^\pm \to K^\pm \pi^+ \pi^-)_{\rm nr,exp}=  $
$14 \pm 6\times 10^{-6}$ and 
${\rm BR}(B^\pm \to K^\pm K^+ K^-)_{\rm nr,exp}= $
$22.5\pm 4.9 \times 10^{-6}$, while the BaBar collaboration
 still has
only the upper limit \cite{fry,BaBar}. 
The inclusion of the  nonresonant contribution in the $B^\pm \to K^\pm
\pi^+ \pi^-$ Dalitz plot analysis \cite{Belle2}
was motivated  by the obvious deficit of the data  in the 
low $K^+\pi^-$ invariant
mass phase space region (see Fig. 11, first row of \cite{Belle2}).
They used  rather simple fit (see Eq. (11) 
\cite{Belle2}) for the nonresonant amplitude. 
Nevertheless, as pointed
out by J. R. Fry \cite{fry}, this contribution is not yet well
understood and more studies of this problem are expected. 
 Calculated ranges for the branching ratios within our model ${\rm BR}(B^\pm \to K^\pm \pi^+
\pi^-)_{\rm nr}=(6.2-12.6)\times 10^{-6}$ and ${\rm BR}(B^\pm \to
K^\pm K^+ K^-)_{\rm nr}= (11-17) \times 10^{-6}\,$ 
 agree with the Belle collaboration's results 
 within 
one standard deviation.
Unfortunately, the experimental statistics is still to low to compare
the distributions of the differential decay rate of the model and the experiment.
It is interesting that our model predicts rather small differential decay width 
distribution in the region of the low $\pi^+ K^-$ invariant mass. 
In order to describe data given in Fig. 11 of \cite{Belle2} 
it seems that one needs such behavior of the 
nonresonant amplitude. 

In addition the results of  \cite{Belle2} indicate existence of 
the
broad structures in the experimental data
at $\sqrt s \simeq 1.3\,$GeV in the $K^+\pi^+\pi^-$ final 
state and at $\sqrt s \simeq 1.5\,$GeV in the $K^+ K^+ K^-$ 
final state.  
Although 
one explanation is that light scalar resonances might be responsible for
this effect  \cite{Belle2}, we suggest that these increases 
might be induced by the nonresonant effects also, what can be seen in the
presented Dalitz plots (Fig. 2).

\begin{figure}
\begin{center}
\includegraphics[width=7cm]{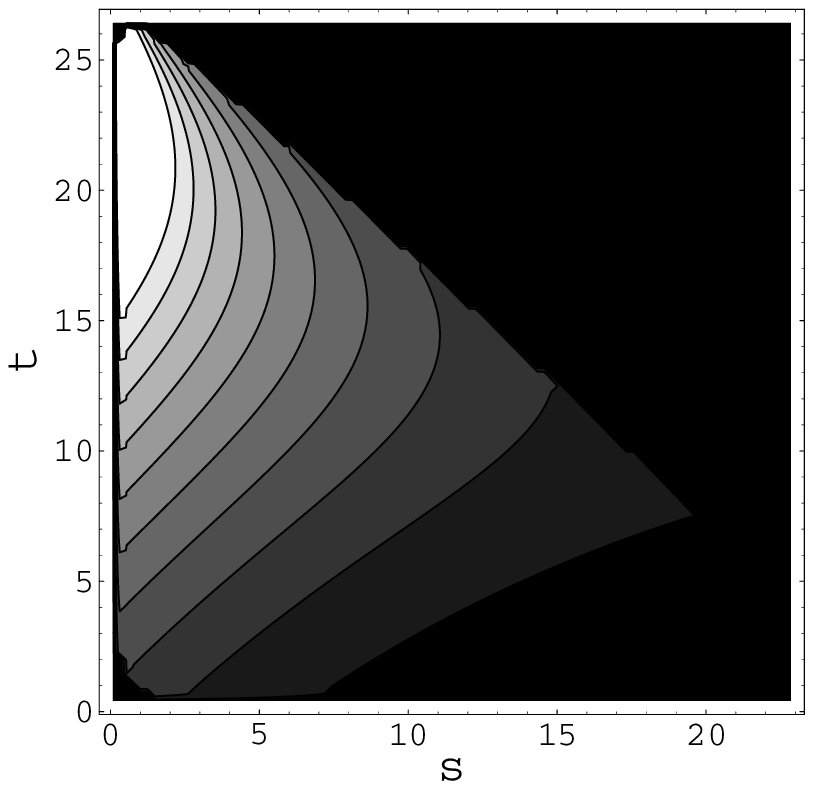}
\includegraphics[width=7cm]{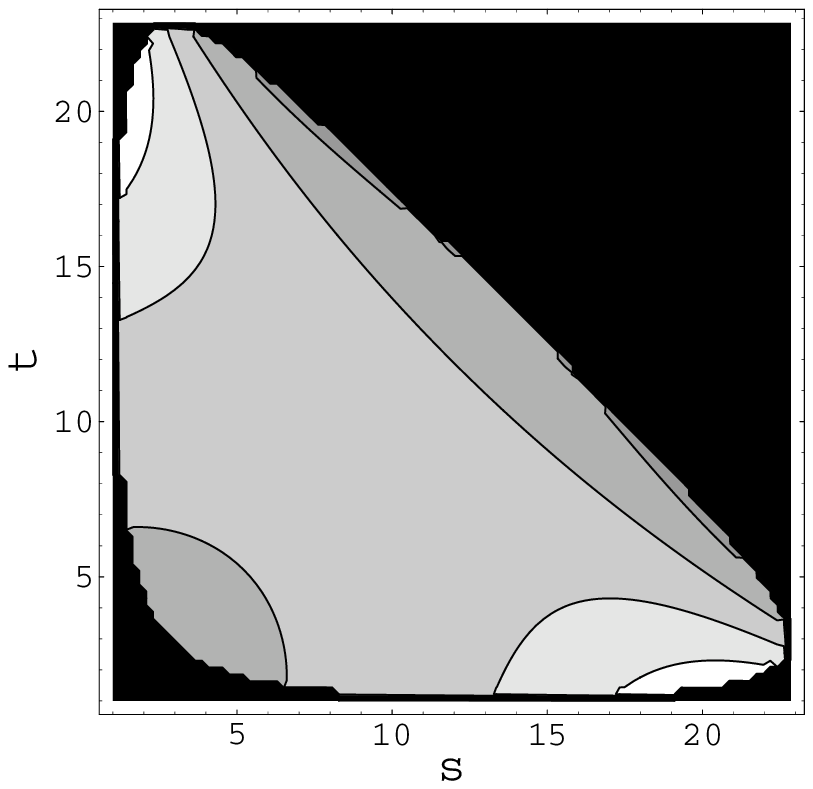}
\label{diagrami2}
\caption{Dalitz plots for the non-resonant $B^- \to K^- \pi^- \pi^+$ (left) 
and $B^- \to K^- K^- K^+$ (right) decay modes.}
\end{center}
\end{figure}

\section{PARTIAL WIDTH ASYMMETRY}

For the resonances in the s-channel, the partial decay width
$\Gamma_p$ for $B^- \to M {\bar M} K^- $, $M = \pi^+$, $K^+$, which
contains both the non-resonant and resonant contributions, is obtained
by integrating the amplitude from $ s_{min} = (m_{R} - 2\Gamma_{R
})^2$ to $ s_{max} = (m_{R} + 2\Gamma_{R })^2$: 
\begin{equation}
\Gamma_p = \frac{1}{(2 \pi)^3} \frac{1}{32 m_B^3}
\int_{s_{min}}^{s_{max}} ds \int_{t_{min}(s)}^{t_{max}(s)} dt~ |{\cal
M}_{nr } + {\cal M}_{r} |^2 .  \label{dwp} 
\end{equation} 
Similarly,
the partial decay width $\Gamma_{\bar p}$ for $B^+ \to M {\bar M}
K^+$, $M = \pi^+$, $K^+$ is defined in a same way. The CP violating
asymmetry is then: 
\begin{equation} A_p = \frac{ |\Gamma_p -
\Gamma_{\bar p}|}{|\Gamma_p + \Gamma_{\bar p}|}.  \label{asym}
\end{equation} 
 It is important to notice that at the phase space region where the invariant mass
of $M^+M^-$ approaches the mass of the ${\cal R}$ resonant state, 
$M^+M^-$ can re-scatter
trough that resonance as it is visualized in Fig. 3 (left figure).
If $Br({\cal R} \to M^+M^-$)
is large, this can lead to a significant absorptive amplitude and 
it contributes to the partial decay width asymmetry. As  
mentioned in Introduction 
and explained in Appendix, such contribution  is exactly 
canceled by the absorptive
part of a resonant decay, where the resonance re-scatters trough the intermediate states equal to
final states (Fig. 3 (right figure)). This implies that
one has  include the factor 
(1 - Br(${\cal R} \to M^+M^-$)) in the equation for the 
partial decay 
asymmetry.

In the calculation of the $\Gamma_p - \Gamma_{\bar p} $,
{ by taking} $V_{ub} = |V_{ub}|e^{i \gamma}$ $ \simeq A \lambda^3
(\bar \rho-i \bar \eta)$, we derive: 
\begin{eqnarray} 
&&\Gamma_p -
\Gamma_{\bar p} = \sin \gamma \frac{4 m_{R}\Gamma_{R}(1 - 
Br({\cal R} \to M^+M^-))}{(2 \pi)^3 32
m_B^3} \nonumber\\ &&\times \int_{s_{min}}^{s_{max}} ds
\int_{t_{min}(s)}^{t_{max}(s)} dt~ \frac{G}{{\sqrt 2}}|V_{ub}|
|V_{us}^*|a_1 <K \pi \pi | O_1| B>_{\rm nr}\nonumber\\ &&\times|{\cal
M} (B^- \to R K^- )| \frac{1}{ (s- m_{R}^2)^2 + (m_{R} \Gamma_{R})^2}
|{\cal M} (R \to \pi^- \pi^+)|,\nonumber\\ \label{asym-}
\end{eqnarray} 
while the $\Gamma_p + \Gamma_{\bar p}$ is given by:
\begin{eqnarray} 
&&\Gamma_p + \Gamma_{\bar p} = 2 \frac{1}{(2 \pi)^3
32 m_B^3} \int_{s_{min}}^{s_{max}} ds \int_{t_{min}(s)}^{t_{max}(s)}
dt~\nonumber\\ &&\times\{ |{\cal M}_{\rm nr}|^2+ |{\cal M} (B^- \to R
K^- ) \frac{1}{ s- m_{R}^2 + i m_{R} \Gamma_{R}} {\cal M} (R \to \pi^-
\pi^+)|^2\}.\nonumber\\ \label{asym+} \end{eqnarray}

The $B^- \to {\cal R} K^- \to K^- M^+ M^-$ amplitudes are obtained
from the experimental data \cite{PDG} and the measured branching
ratios for $B^- \to {\cal R} K^- $ and $ R \to M^+ M^-$ are given in
Table 1.

\begin{table}
\begin{center}
\begin{tabular}{|c|c|c|c|} \hline
 {\cal R}       & $B^- \to  {\cal R} K^-$  & $  {\cal R} \to \pi^+ \pi^- $ & 
 ${\cal R} \to K^+ K^-$ \\ \hline
 $\phi$        &  $(7.9 \pm 2.0)\times 10^{-6}$ & & $(49.2 \pm 0.7)$\% \\ 
 $J/\psi$       & $(1.01\pm 0.05)\times 10^{-3} $ & 
 $(1.47\pm 0.23)\times 10^{-4} $  
                & $(2.37\pm 0.0.31)\times 10^{-4} $ \\  
 $\chi_{c0}$    & $(6.5 \pm 1.1)\times 10^{-4} $  & 
 $(5.0 \pm 0.7)\times 10^{-3} $ & $(5.9 \pm 0.9) \times 10^{-3}$ \\  
 $\chi_{c1}$    & $(6.0 \pm 2.4)\times 10^{-4} $  & 
 $ <2.1 \times 10^{-3}$ & $ <2.1 \times 10^{-3}$ \\  
 $\psi(2S)$     & $(6.6 \pm 0.6)\times 10^{-4} $  & $(8 \pm 5)\times 10^{-5} $ 
                & $(1.0 \pm 0.7)\times 10^{-4}$ \\  \hline 
\end{tabular}
\end{center}
\caption{The decay $B^- \to  {\cal R} K^-$ width and the branching ratios for 
${\cal R} \to M^+ M^-$. }
\end{table}

For the scalar resonance exchange ($\chi_0$ in our case) in the
$B^- \to S K^- \to M^+ M^- K^-$ decay, we have:
\begin{equation}
{\cal M}(B^- \to S P_1(q_1) \to P_1(q_1)P_2(q_2)P_3(q_3))= $$$$ {\cal
M}(B^- \to S P_1(q_1))\frac{1}{m_{23}^2-m_S^2+i\Gamma_S m_S}{\cal M}(S
\to P_2(q_2)P_3(q_3) )\,,
\end{equation}
where $m^2_{23}=(q_2+q_3)^2$ while $m_S$ and $\Gamma_S$ are the mass
and the decay width of the scalar resonance respectively. We find
${\cal M}(B^- \to \chi_{c0} K^-)=3.34 \times 10^{-7}$ GeV, ${\cal
M}(\chi_{co}\to \pi^+\pi^-)=0.118$ GeV and ${\cal M}(\chi_{co}\to K^+
K^-)=0.132$ GeV.

The amplitude for the $B^-$ decay into light vector and pseudoscalar
resonance and the amplitude for the vector meson decay into two
pseudoscalar states are given by:
\begin{equation}
{\cal M}(B^- \to V(\varepsilon) P_1(q_1))=K q_1 \cdot \varepsilon^*\,,
\qquad
{\cal M}(V \to P_2(q_2) P_3(q_3))=\frac{g_{VP_2P_3}}{\sqrt 2} (q_2 -
q_3)\cdot \varepsilon.
\end{equation}
The amplitude for the three-body resonant decay for this case is:
\begin{equation}
{\cal M}(B^- \to V(\varepsilon) P_1(q_1) \to P_1(q_1)P_2(q_2)
P_3(q_3))= $$$$\frac{K g_{VP_2P_3}}{\sqrt 2} \frac{-q_1 \cdot (q_2
-q_3)+
(q_1\cdot(q_2+q_3)(m^2_2-m^2_3))/m_V^2}{m^2_{23}-m_V^2+i\Gamma_V
m_V}\,,
\end{equation}
where $m^2_{23}=(q_2+q_3)^2$, while $m_1$, $m_2$, $m_3$ and $m_V$ are
the masses of particles $P_1$, $P_2$, $P_3$ and $V$ respectively and
$\Gamma_V$ is the width of the vector resonance. Using above formulas,
we find the expression for the resonance exchange in the $s-$channel:
\begin{equation}
{\cal M}(B^- \to V K^- \to K^- M^+ M^-)= 
\frac{K_{V} g_{VMM}}{2\sqrt 2} 
\frac{m_B^2+2m_M^2+m_K^2-2t-s}{s-m_{V}^2+i\Gamma_{V} m_{V}}\,,
\end{equation}
where $M$ stands for $K$ or $\pi$. In the case of the $K^-K^+ K^\pm$
mode the contributions coming from the $s$ and $t$ channels are
completely symmetric. Values of $K_V$ and $g_{VP_1P_2}$ are given in
Table 2.

\begin{table}
\begin{center}
\begin{tabular}{|c|c|c|c|} \hline
 $V$            & $K_V$ & $g_{V\pi\pi}$ & $g_{VKK} (g_{VK\pi})$  \\ \hline 
 $\phi$         & $2.26 \times 10^{-9}$ &  & 6.34 \\  
 $J/\psi$       & $1.41 \times 10^{-7}$ & $2.76 \times 10^{-4}$ & 
 $4.37 \times 10^{-3}$  \\ 
 $\psi(2S)$     & $2.04 \times 10^{-7}$ & $1.41 \times 10^{-3}$ & 
 0.166  \\ 
 $\chi_{c1}$    & $1.68 \times 10^{-7}$ & $<0.0126$ & $<0.0134$  \\ \hline    
\end{tabular}
\caption{The  parameters used in our numerical calculations.}
\end{center}
\end{table}

\begin{table}
\begin{center}
\begin{tabular}{|c|c|c|c|c|c|} \hline
& $\bar\rho=0.118$ & $\bar\rho=0.118$
 & $\bar\rho=0.273$ & $\bar\rho=0.273$ & $\bar\rho=0.222$
\\ 
 & $\bar\eta=0.305$ & $\bar\eta=0.393$
 & $\bar\eta=0.305$ & $\bar\eta=0.393$
 & $\bar\eta=0.339$ \\ \hline
$A_p$($\psi(2S)$) & 10.2\%  & 13.0\% & 10.3\%  & 13.1\% & 11.3\%\\ 
$A_p$($J/\psi$)   & 0.8\% & 1.1\%   & 0.8\% & 1.1\%  & 0.9\%\\ 
$A_p$($\chi_{c1}$) & 3.5\%  & 4.5\% & 3.5\%  & 4.5\% & 3.9\% \\ 
$A_p$($\chi_{c0}$) & 17.3\%  & 21.8\%  & 17.6\%  & 22.1\%  & 19.3\%\\ \hline
\end{tabular}
\caption{The  partial width asymmetry for 
$B^- \to K^- \pi^+ \pi^-$, calculated with $g =0.56$ and given 
$\bar\rho$ and $\bar\eta$.  
 $A_p$($\chi_{c1}$) is obtained by taking the upper bound for 
 $g_{VMM}$.} 
\end{center}
\end{table}

\begin{table}
\begin{center}
\begin{tabular}{|c|c|c|c|c|c|} \hline
 &$\bar\rho=0.118$ & $\bar\rho=0.118$
 & $\bar\rho=0.273$ & $\bar\rho=0.273$ & $\bar\rho=0.222$
\\ 
 & $\bar\eta=0.305$ & $\bar\eta=0.393$
 & $\bar\eta=0.305$ & $\bar\eta=0.393$
 & $\bar\eta=0.339$ \\ \hline
$A_p$($\psi(2S)$) & 13.5\%  & 17.3\% & 13.7\%  & 17.3\% & 15.1\%\\ 
$A_p$($J/\psi$)   & 1.2\% & 1.6\%   & 1.2\% & 1.6\%  & 1.4\%\\ 
$A_p$($\chi_{c1}$) & 5.0\%  & 6.4\% & 5.0\%  & 6.5\% & 5.6\% \\ 
$A_p$($\chi_{c0}$) & 12.8\%  & 16.1\%  & 12.9\%  & 16.3\%  & 14.2\%\\ \hline
\end{tabular}
\caption{The  partial width asymmetry for 
$B^- \to K^- \pi^+ \pi^-$, calculated with $g =0.27$ and given 
$\bar\rho$ and $\bar\eta$. $A_p$($\chi_{c1}$) is obtained by 
taking the upper bound for $g_{VMM}$.}
\end{center}
\end{table}

\begin{table}
\begin{center}
\begin{tabular}{|c|c|c|c|c|c|} \hline
 & $\bar\rho=0.118$ & $\bar\rho=0.118$
 & $\bar\rho=0.273$ & $\bar\rho=0.273$& $\bar\rho=0.222$
\\ 
 & $\bar\eta=0.305$ & $\bar\eta=0.393$
 & $\bar\eta=0.305$ & $\bar\eta=0.393$ & $\bar\eta=0.339$
\\ \hline
$A_p$($\phi$)      & 0.3\% & 0.3\%  & 0.3\% & 0.3\%  & 0.3\% \\ 
$A_p$($\psi(2S)$) & 3.1\%  & 3.8\%  & 3.0\% & 3.7\% & 3.3\%\\ 
$A_p$($J/\psi$)   & 0.03\% & 0.04\% & 0.03\% & 0.04\% & 0.03\%\\ 
$A_p$($\chi_{c1}$) & 0.5\%  & 0.7\% & 0.5\%  & 0.3\% & 0.6\%\\ 
$A_p$($\chi_{c0}$) & 28.8\%  & 35\%  & 27.6\%  & 33.8\%  & 30.6\%\\ \hline
\end{tabular}
\caption{The partial width asymmetry for 
$B^- \to K^- K^+ K^-$, calculated with $g =0.56$ and given 
$\bar\rho$ and $\bar\eta$. 
$A_p$($\chi_{c1}$) is obtained by taking the upper bound for 
$g_{VMM}$.}
\end{center}
\end{table}

\begin{table}
\begin{center}
\begin{tabular}{|c|c|c|c|c|c|} \hline
& $\bar\rho=0.118$ & $\bar\rho=0.118$
 & $\bar\rho=0.273$ & $\bar\rho=0.273$& $\bar\rho=0.222$
\\ 
 & $\bar\eta=0.305$ & $\bar\eta=0.393$
 & $\bar\eta=0.305$ & $\bar\eta=0.393$& $\bar\rho=0.339$
\\ \hline
$A_p$($\phi$)      & 0.3\% & 0.3\% & 0.3\% & 0.3\% & 0.3\% \\ 
$A_p$($\psi(2S)$) & 8.1\%  & 10.1\% & 7.9\% & 9.8\% & 8.8\% \\ 
$A_p$($J/\psi$)   & 0.55\%  & 0.71\% & 0.55\% & 0.71\% & 0.61\% \\ 
$A_p$($\chi_{c1}$) & 3.0\%  & 3.8\%  & 3.0\%  & 3.8\%  & 3.3\% \\ 
$A_p$($\chi_{c0}$) & 23.1\% & 28.7\% & 22.5\% & 28.0\% & 25\%\\ \hline
\end{tabular}
\caption{The  partial width asymmetry for 
$B^- \to K^- K^+ K^-$, calculated with $g =0.27$ and given $\bar\rho$
and $\bar\eta$. $A_p$($\chi_{c1}$) is obtained by taking the upper
bound for $g_{VMM}$.}
\end{center}
\end{table}

The results for the asymmetries are presented in Tables 3-6.  Tables 3
and 5 contain the asymmetries for $g=0.56$. The off-shell mass effects
might reduce this coupling as mentioned in \cite{FOP}, and therefore
we present the partial width asymmetries for $g=0.27$ (Tables 4 and
6). We calculate asymmetries for the ranges $\bar \rho=0.118-0.273$
(the average value 0.222) and $\bar \eta=0.305-0.393$ (the average
value 0.339) as in \cite{ckmfit}. 
The subtraction of $Br({\cal R} \to M^+M^-)$ in Eq. 
(\ref{asym-}) makes a sizable effect in the case of the
$B^- \to K^- \phi \to M^+ M^- K^-$ asymmetry, but it is 
negligible in the case of partial width asymmetry 
in the neighborhood  of charmonium resonances.  
Then we can draw the following
conclusions: In the case of $B^- \to K^- \pi^+ \pi^-$, all partial
width asymmetries are not very large. The largest asymmetry was found
in the case of $\chi_{c0}$ resonance and then in the case of
$\psi(2S)$. The partial width asymmetry $A_p(\chi_{c1})$ is calculated
by taking the upper bounds for $\chi_{c1} M^+ M^-$ coupling.  All 
these asymmetries are rather stable 
on the variations of $g$.  In the case
of $B^- \to K^- K^+ K^-$ the situation is different. Calculated 
partial width asymmetries
except the $A_p(\chi_{c0})$ are smaller than in the case of $B^- \to
K^- \pi^+ \pi^-$. They depend more on the variations of the $g$
coupling.  The only relatively sizable partial width 
asymmetry in addition to
$A_p(\chi_{c0})$ is $A_p(\psi(2s))$.

We have also estimated the partial width asymmetry for the $B^-\to
\chi_{c2} K^-$ channel, by assuming the $B_{\chi_{c2} K}$ coupling to
be of the same size as for the vector (scalar) mesons and we found it
negligible.

\section {SUMMARY}

In this paper we have investigated the partial width asymmetry for the
$B^- \to M^+ M^- K^-$, $M= \pi, K$ decays which results from the
interference of non-resonant and resonant amplitudes.

First, we have calculated the non-resonant branching ratios and found
that the model we use gives the decay rates in the reasonable
agreement with the Belle collaboration results
\cite{Belle2}. Comparing the Dalitz plots for the non-resonant decay
modes obtained from our model with the experimental data
\cite{Belle2}, we find that our model reproduces the data quite
well. The inclusion of the $B^{*}_{0}$ scalar meson is rather
insignificant, contributing only by few percents to the rate.

We then consider the partial width asymmetries for a few resonant
decay modes for which the amplitude does not contain the weak phase
$\gamma$.  In the case of $B^- \to \pi^+ \pi^- K^-$ the largest
partial width asymmetry arises from the interference of the
non-resonant amplitude with the resonant amplitude coming from the
$\chi_{c0}$ and $\psi(2S)$ states.  In the case of $B^- \to K^- K^+
K^-$ the largest partial width asymmetry comes from the $\chi_{c0}$
scalar resonance, while and is about $~10\%$ in the case of $\psi(2S)$
state.

\vspace{0.5cm}
{\bf ACKNOWLEDGMENTS}
\vspace{0.5cm}
 
We thank our colleagues B. Golob and P. Kri\v zan for stimulating
discussions on experimental aspects of this investigation and D. Be\'
cirevi\' c for very useful comments. The research of S. F. and
A. P. was supported in part by the Ministry of Education, Science and
Sport of the Republic of Slovenia.
\newpage
\vspace{0.5cm}
{\bf \large APPENDIX}
\vspace{0.5cm}

 \begin{figure}
\begin{center}
\includegraphics[width=15cm]{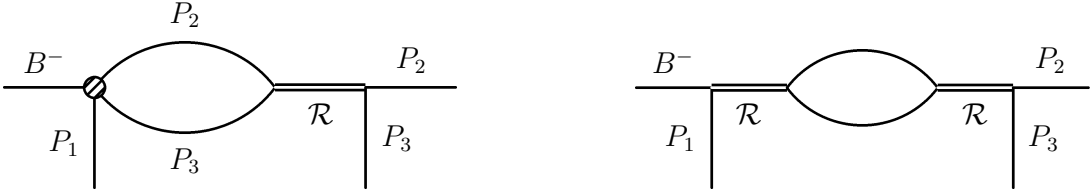}
\caption{Diagrams presenting the non-resonant
(left) and the resonant (right) contributions to dispersive part of
the amplitude in the phase space region of the $P_2$ and $P_3$
invariant mass close to the ${\cal R}$ mass.  Blob in the left diagram
presents the non-resonant weak decay mode (see Fig. 1).}
\end{center}
\label{FSI}
\end{figure}

Following the approach of \cite{SEW},
 the total amplitude contributing to the partial decay width 
 $\Gamma_p$
can be written as a sum of the resonant and nonresonant contributions as defined in
Eq. (21)  in the following form:
\begin{equation}
{\cal M}={\cal M}_{nr}+{\cal M}_{r}=Te^{-i\gamma}+P+R\,,
\end{equation}
where $T$ is the nonresonant tree contribution, $P$ the
nonresonant penguin contribution and $R$ the resonant
contribution to the amplitude. The partial width asymmetry defined in
Eq. (22) is proportional to:
\begin{equation}
A_p \propto \sin\gamma \;\Im(T(P^*+R^*))\,,
\end{equation}
where $\Im(A)$ stands for the imaginary part of $A$ (similarly
$\Re(A)$ stands for the real part of $A$).  If we neglect the small
imaginary part of the penguin Wilson coefficients, $T$ and $P$ will
have the same strong phase. This implies that the only contribution to the partial
decay asymmetry will come from the interference of the tree
nonresonant and the resonant amplitude.  One can write:
\begin{equation}
A_p \propto \Im(T)\Re(R)-\Im(R)\Re(T)\,.
\label{A0}
\end{equation}
The imaginary part of $T$ is given by the absorptive part of the left
diagram on the Fig. 3. Using Cutkosky's rules, its
contribution can be written as:
\begin{equation}
\Im(T)\Re(R)=\frac{(2\pi)^4}{2}\int \Re(T) \Re(R)\Re(S) d\Phi\,,
\label{A1}
\end{equation}
where the integration is taken over the $P_2 P_3$ phase space.  
Here $S$
denotes the strong re-scattering amplitude of $P_2 P_3$ trough the
resonance ${\cal R}$ visualized in Fig. 4.  Similarly, the
imaginary part of $R$ is given by the absorptive part of the right
diagram on the Fig. 3, where now the sum of all possible
intermediate states into which ${\cal R}$ decays should be taken into
account. One can  separate this contribution into the part with $P_2$
and $P_3$ as an intermediate state ($\Im(R)_{P_{2,3}}$) and the part
with all other intermediate states ($\Im(R)^\prime$). Again with the use 
of 
Cutkosky's rules, one obtains:
\begin{equation}
\Im(R_{P_{2,3}})\Re({T}=
\frac{(2\pi)^4}{2}\int \Re(T) \Re(R) \Re(S) d\Phi\,.
\label{A2}
\end{equation}
The right hand sides of (\ref{A2}) and (\ref{A1}) are
equal and therefore this two contributions to (\ref{A0}) cancel among
themselves and we have:
\begin{equation}
A_p \sim \Im(R)^\prime\Re(T) \sim \Gamma_R(1-Br({\cal R}\to P_2 P_3)
\Re(T)\,.
\end{equation}
This cancellation is obviously 
a result of the unitarity and it maintains the equality of the 
total decay widths for the meson and the anti-meson 
as required by CPT theorem \cite{GHW}. 
That was already noticed by \cite{EGM} and \cite{SEW}, where
the more general proof is presented.

\begin{figure}
\begin{center}
\includegraphics[width=3cm]{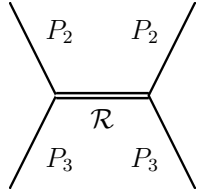}
\label{RFSI2}
\caption{Re-scattering of the $P_2$ and $P_3$ states 
trough the resonance ${\cal R}$.}
\end{center}
\end{figure}

\end{document}